\documentclass[11pt,leqno]{article}

\usepackage[english]{babel}
\usepackage[utf8]{inputenc}

\usepackage{a4wide}
\usepackage{xcolor}

\usepackage{tikz}
\usetikzlibrary{arrows,automata,decorations.pathreplacing, shapes.multipart}

\usepackage{subcaption}

\makeatletter
\tikzset{circle split part fill/.style  args={#1,#2}{%
 alias=tmp@name, 
  postaction={%
    insert path={
     \pgfextra{%
     \pgfpointdiff{\pgfpointanchor{\pgf@node@name}{center}}%
                  {\pgfpointanchor{\pgf@node@name}{east}}%
     \pgfmathsetmacro\insiderad{\pgf@x}
      \fill[#1] (\pgf@node@name.base) ([xshift=-\pgflinewidth]\pgf@node@name.east) arc
                          (0:180:\insiderad-\pgflinewidth)--cycle;
      \fill[#2] (\pgf@node@name.base) ([xshift=\pgflinewidth]\pgf@node@name.west)  arc
                           (180:360:\insiderad-\pgflinewidth)--cycle;            
         }}}}}  
 \makeatother  

\tikzset{state_left/.style={circle split part fill={white!100,black!80},rotate=270}}

\tikzset{state_right/.style={circle split part fill={white!100,black!80},rotate=90}}

\usepackage{amsmath}
\usepackage{amssymb} 
\usepackage{amsthm}
\usepackage{amsfonts}
\usepackage{listings}
\usepackage{url}

\usepackage{graphicx}
\usepackage[colorlinks=true, allcolors=blue]{hyperref}

\usepackage{dirtytalk}

\newtheorem*{defn}{Definition}
\newtheorem{teo}{Theorem}
\newtheorem{lema}{Lemma}[section]
\newtheorem{prop}{Proposition}[section]
\newtheorem{coro}{Corollary}[section]
\newtheorem*{nota}{Notation}

\newcommand{\enne}[0]{\mathbb{N}}

\newcommand{\emptyword}[0]{\varepsilon}

\newcommand{\conj}[1]{\{#1\}}

\newcommand{\abu}{\mathcal{A}}

\newcommand{\ele}{\mathcal{L}}
\newcommand{\eme}{\mathcal{M}}

\newcommand{\ese}{\mathcal{S}}

\newcommand{\ubu}{U}

\newcommand{\rat}{\text{Rat}}

\newcommand{\drat}{\text{DRat}}

\newcommand{\autom}{(Q,M,E,I,T)}
\newcommand{\shusplit}{(Q,A,E,I,T)}

\newcommand{\emptywords}{((\emptyword,\emptyword),(\emptyword,\emptyword))}

\title{On Shuffling and Splitting Automata}
\author{Ignacio Mollo Cunningham\footnote{Contact: \href{mailto:imcgham@gmail.com}{imcgham@gmail.com}}\\
\footnotesize{Instituto de Ciencias de la Computación}\\
\footnotesize{Universidad de Buenos Aires}\\
\footnotesize{Buenos Aires, Argentina}}

\begin{document}
\maketitle

\begin{abstract}

We consider a class of finite state three-tape transducers which models the operation of shuffling and splitting words. We present them as automata over the so-called Shuffling Monoid. These automata can be seen as either shufflers or splitters interchangeably. We prove that functionality is decidable for splitters, and we also show that the equivalence between functional splitters is decidable. Moreover, in the deterministic case, the algorithm for equivalence is polynomial on the number of states of the splitter.\\

\end{abstract}

\tableofcontents

\section*{Introduction}

In this work we study automata which model the operations of shuffling and splitting words. We define them as automata over a special monoid $U$ which we call the Shuffling Monoid, and we call them \textit{spliffers}. This is a portmanteau between the words \textit{shuffler} and \textit{splitter}, which is appropriate given that these machines can be thought to perform either of these two operations depending of the point of view. We also analyze the deterministic variant of these spliffers. 
These were used by Álvarez, Becher and Carton in \cite{alvarez2019finite} to characterize the notion of independence between normal numbers. 

Similar kinds of automata have been considered before. For instance, in \cite{boasson2015rational} Boasson and Carton study the case of \textit{selectors}, which are two-tape transducers which model the operation of selecting subwords of a word. This is very similar to our case, because the act of splitting words can be thought as selecting two different interwoven subwords of a word. However, the additional rigidity of the relations defined by spliffers make them interesting on their own.

In this work we consider the rational relations defined by spliffers and deterministic spliffers. We analyze the closure properties of their respective rational relations, and consider the equivalence problem for them. We prove that equivalence between spliffers is decidable under an additional condition of \textit{functionality}: we do this by adapting the technique introduced by B{\'e}al, Carton, Prieur and Sakarovitch in \cite{beal2003squaring} to prove the decidability of the equivalence problem of functional transducers. Furthermore, all deterministic spliffers are also functional, and in this case the algorithm can be performed in polynomial time over the number of states of the spliffer. 

It's worth noting that the question of decidability of equivalence of deterministic spliffers is already answered positively by the work of Harju and Karhum{\"a}ki in \cite{harju1991equivalence}. This is because the Shuffling Monoid over which our spliffers are defined is graded. However, our algorithm is polynomial in the deterministic case and applies to a larger family of spliffers than merely the deterministic ones.

\section{The Shuffling Monoid and its Automata}\label{overview}

We fix an alphabet $A$, and we consider the monoid $A^\star\times A^\star\times A^\star$ where the product is point-wise concatenation. We denote by $\emptyword$ the empty word in $A^\star$.

\subsection{Shuffling and Splitting Automata}

\begin{defn}[The Shuffling Monoid]
We define the Shuffling Monoid $\ubu$ to be the sub-monoid of $A^\star\times A^\star\times A^\star$ generated by the finite set $$G = \conj{(a,\emptyword,a):a\in A}\cup\conj{(\emptyword,a,a):a\in A}.$$
\end{defn}

Intuitively, $\ubu$ is the set of triples $(l,r,s)\in M$ such that $s$ can be obtained by interlacing $l$ and $r$. That is, $s$ can be written as $s=r_1l_1r_2l_2\dots r_kl_k$ where $l_i,r_j\in A\cup\conj{\emptyword}$ for all $i,j$ and $l_1\dots l_k=l$, $r=r_1\dots r_k$. 

$U$ is an example of a graded monoid, with a length function $\varphi:U\rightarrow\enne$ given by $\varphi(l,r,s) = |s|$. In particular, $U$ is finitely decomposable, that is, every element $(l,r,s)\in U$ admits only a finite number of ways to be written as a product of generating elements in $G$. Quite frequently, this number of possible decompositions is greater than one: for instance $(w,w,ww)$ can be written as either $(w,\emptyword,w)(\emptyword,w,w)$ or $(\emptyword,w,w)(w,\emptyword,w)$ whatever the choice of word $w\in A^+$. Moreover, this last example can occur as part of a longer tuple, leading to situations when it's ambiguous from which of the two first coordinates a particular letter is taken to be shuffled. 

As it can be done with any monoid, we define the class of automata with transitions labelled by elements of $U$.

\begin{defn}[Spliffer]
A spliffer over the alphabet $A$ is a finite automaton over $\ubu$, with its transitions labeled by elements in $G$. We usually note $\ese = \shusplit$, where $Q$ is a set of states, $E$ the set of transitions, $I\subseteq Q$ the set of initial states, and $T\subseteq Q$ the set of final states.
\end{defn}

As it is customarily done with automata over general monoids, we consider \textit{computations} on them, that is, paths over their graphs of transitions. These paths have labels in $\ubu$, consisting on the product of the labels of all edges in the path. We call a computation \textit{successful} if it starts in a state from $I$ and ends in a state from $T$. Then, a spliffer $\ese$ \textit{accepts} a subset of $\ubu$, comprised by the labels of every successful run. We denote this subset by $|\ese|$, and we call it the \textit{behaviour} of $\ese$. 

The set of behaviors of spliffers corresponds exactly with the set of rational sets of $\ubu$, noted as $\rat(\ubu)$. This is because $\ubu$ is finitely generated: so, trivially, any automata over $\ubu$ is equivalent to an automata over $\ubu$ labeled by elements of the generating set $G$. This is exactly what we defined as spliffers.

\begin{figure}[h]
        \centering
    \begin{tikzpicture}[->,initial text=,semithick,auto,inner sep=3pt]
      \tikzstyle{every state}=[minimum size=17];
      \node[state,initial left] (qi) at (0,0){};
      \node[state] (p1) at (3,-1){};
      \node[state] (p2) at (6,-1){};
      \node[state]  (q1) at (3,1){};
      \node[state]  (q2) at (6,1){};
      \node[state, accepting right] (qf) at (9,0){};
      \path (qi) edge node {$(a,\emptyword|a)$} (q1);
      \path (q1) edge node {$(\emptyword,b|b)$} (q2);
      \path (q2) edge node {$(a,\emptyword|a)$} (qf);
      \path (qi) edge node {$(\emptyword,a|a)$} (p1);
      \path (p1) edge node {$(\emptyword,b|b)$} (p2);
      \path (p2) edge node {$(a,\emptyword|a)$} (qf);
    \end{tikzpicture}
    \caption{A simple spliffer accepting the set $\conj{(aa,b,aba),(a,ab,aba)}$}
    \label{fig:elementary-shu-spl}
\end{figure}
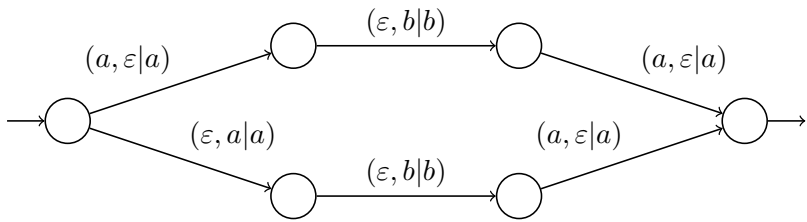

Figure \ref{fig:elementary-shu-spl} depicts an example of this kind of automata. Note that we usually write the first two coordinates of each label separately from the third.

Informally,  we may think of a spliffer as an automaton which reads from three tapes. Depending on which of the tapes are considered \say{input} or \say{output}, we get either a machine that shuffles two input tapes together, or a machine that splits the input apart into two. Sometimes it is useful to explicitly distinguish these two operational modes, and we do so by virtue of the following notations.

\begin{nota}
\begin{itemize}

\item[]

\item A \upshape{shuffler} \itshape is a spliffer for which the first two coordinates of an element $(l,r,m)$ of $\ubu$ are considered \say{input}, and the third is considered \say{output}. Intuitively, a shuffler is a three-tape automaton (two inputs, one output), such that the inputs can be read independently from one another and are repeated to the output. 

\item A \upshape{splitter} \itshape is a spliffer for which the first two coordinates of an element $(l,r,m)$ of $\ubu$ are considered \say{output}, and the third is considered \say{input}. Intuitively, a splitter is a three-tape automaton (two outputs, one input), that takes an input and splits it into two.
\end{itemize}
\end{nota}

The automaton in Figure \ref{fig:elementary-shu-spl}, for instance, can be thought to be splitting the string $aba$ in two different ways, or producing the same string by shuffling two different pairs of strings. Effectively, and without a notion of time, shuffling and splitting are indistinguishable. In the following, we talk about the shuffler, or the splitter, associated with a spliffer $\ese$ to mean that we are regarding $\ese$ in one of these two ways. In most cases we consider, it's more natural to work with splitters than shufflers.

We are also interested in studying a deterministic variant of these machines. We start by defining the notion of deterministic splitter. We do it this way because the input of a splitter is simpler than the input of a shuffler.

\begin{defn}[Deterministic splitter]
We say that a splitter $S$ is deterministic if the following three conditions hold:  
\begin{itemize}
    \item There is a single initial state $i$ in $S$. 
    \item For every input $s\in A$ and every state $q$, there is at most one transition $q\xrightarrow{(l,r,s)}q'$ with $(l,r,s)\in G$. 
    \item For every state $q$, all transitions leaving from $q$ write to the same tape. That is, if $q\xrightarrow{(l',r',s')}q'$ and $q\xrightarrow{(l'',r'',s'')}q''$ are two different transitions in $S$, then $l' = l'' = \emptyword$ or $r' = r'' = \emptyword$. 
\end{itemize}
\end{defn}

These conditions ensure that there's only one thing a deterministic splitter can do for every input. This definition provides us with a way to think about deterministic spliffers as well.  

\begin{defn}
    We say that a spliffer $S$ is deterministic if its associated splitter is deterministic.
\end{defn}

In particular:

\begin{lema}\label{lemma:det_spliffers_are_unambiguous}
    Deterministic spliffers are unambiguous.
\end{lema}

...something which is common sense to ask when defining deterministic machines. 

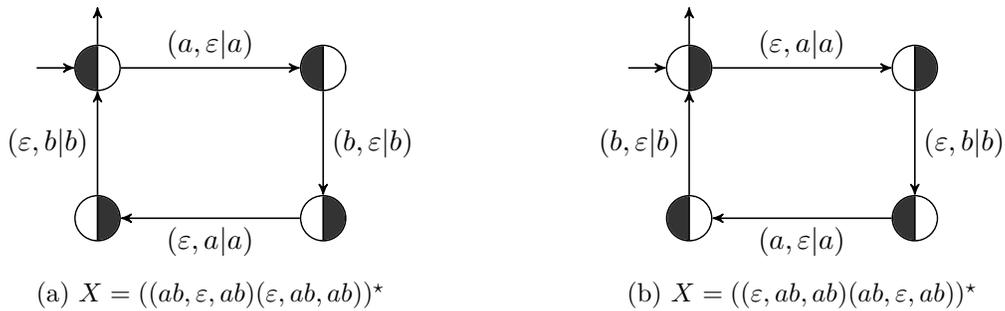
\begin{figure}[h]
    \begin{subfigure}[b]{0.5\textwidth}
    \centering
    \begin{tikzpicture}[->,>=stealth',initial text=,semithick,auto,inner sep=3pt]
      \tikzstyle{every state}=[shape=circle split, minimum size=17];
      \node[state, style=state_left] (p0) at (0,0){};
      \node[state, style=state_left] (p1) at (3,0){};
      \node[state, style=state_right] (p2) at (3,-2){};
      \node[state, style=state_right] (p3) at (0,-2){};
      \draw[->] (p0.west) -- ++(0,0.5cm);
      \draw[<-] (p0.south) -- ++(-0.5,0cm);
      \path (p0) edge node {$(a,\emptyword|a)$} (p1);
      \path (p1) edge node {$(b,\emptyword|b)$} (p2);
      \path (p2) edge node {$(\emptyword,a|a)$} (p3);
      \path (p3) edge node {$(\emptyword,b|b)$} (p0);
    \end{tikzpicture}
    \subcaption{$X=((ab,\emptyword,ab)(\emptyword,ab,ab))^\star$}
    \end{subfigure}
    \begin{subfigure}[b]{0.5\textwidth}
         \centering
         \begin{tikzpicture}[->,>=stealth',initial text=,semithick,auto,inner sep=3pt]
      \tikzstyle{every state}=[shape=circle split, minimum size=17];
      \node[state, style=state_right] (p0) at (0,0){};
      \node[state, style=state_right] (p1) at (3,0){};
      \node[state, style=state_left] (p2) at (3,-2){};
      \node[state, style=state_left] (p3) at (0,-2){};      
      \draw[->] (p0.east) -- ++(0,0.5cm);
      \draw[<-] (p0.north) -- ++(-0.5,0cm);
      \path (p0) edge node {$(\emptyword,a|a)$} (p1);
      \path (p1) edge node {$(\emptyword,b|b)$} (p2);
      \path (p2) edge node {$(a,\emptyword|a)$} (p3);
      \path (p3) edge node {$(b,\emptyword|b)$} (p0);
    \end{tikzpicture}
    \subcaption{$X=((\emptyword,ab,ab)(ab,\emptyword,ab))^\star$}
    \end{subfigure}
    \caption{Two different deterministic Spliffers accepting the same set}
    \label{fig:elementary-det-shu-spl}
\end{figure}

An element in $\ubu$ may admit (and most of them do) more than one decomposition as a product of elements in $G$. For instance, $(ab,ab,abab)$ can be written in the following two ways: 
$$
(a,\emptyword,a)(b,\emptyword,b)\cdot(\emptyword,a,a)(\emptyword,b,b) = (\emptyword,a,a)(\emptyword,b,b)\cdot (a,\emptyword,a)(b,\emptyword,b). 
$$ 
However, a deterministic spliffer that accepts it will do so in only one way. As an example, consider the set $X=\conj{((ab)^n,(ab)^n,(ab)^{2n}): n\geq1}\subseteq\ubu$. Figure \ref{fig:elementary-det-shu-spl} depicts two deterministic spliffers that accept $X$ in different ways. The states are colored that way so that it is easy to tell at a glance from which tape they read (or to which tape they write).

If one ignores the two output tapes in a splitter, and considers only the projections of the transitions on the input, one obtains a classical automaton over $A^\star$. We call this the \textit{underlying input automaton} of a splitter. Similarly, via projection on the output, we can define the \textit{underlying output transducer} of a splitter.

These are immediate corollaries from the previous definitions:

\begin{coro}

\begin{itemize}
    \item[]
    
    \item The underlying input automaton of a deterministic splitter is a classical deterministic automaton over $A^\star$.

    \item The underlying output transducer of a deterministic splitter is a deterministic two-tape transducer over $A^\star\times A^\star$. 
\end{itemize}

\end{coro}

\begin{coro}\label{projection-is-regular}
The projection onto the third coordinate of the behavior $|\abu|$ of a spliffer $\abu$ is a rational language. 
\end{coro}

This last property follows from the fact that the projection $\pi_3:M\rightarrow A^\star$ onto the third coordinate is a monoid morphism. This property, as well as those that follow, will help us with identifying subsets of $\ubu$ that are \textit{not} the behavior of spliffers.

\begin{defn}
    Let $S$ be a splitter. 
    \begin{itemize}
        \item We say that $S$ is functional if the behavior of $S$ is a partial function. That is, whenever $(u_1,v_1,s),(u_2,v_2,s)\in |S|$ for some word $s\in A^\star$ then $u_1=u_2$ and $v_1 = v_2$.
        \item We say that $S$ is injective if the behavior of $S$ is an injection. That is, whenever $(u,v,s_1),(u,v,s_2)\in |S|$ for words $u,v\in A^\star$ then $s_1=s_2$. 
    \end{itemize}
\end{defn}

\begin{prop}\label{behavior-is-injective-function}
    Deterministic splitters are functional and injective. That is, their behavior is an injective function.
\end{prop}

\begin{proof}
    We start by showing that $\ese$ is injective.
    
    Let $r_1 = (u,v,s_1)$ and $r_2 = (u,v,s_2)$ be two elements of $|\ese|$. We'll show that $s_1$ and $s_2$ must be equal. As they are accepted by $\ese$, for $i=1,2$, there must exist decompositions:
    $$ 
    r_i = (u_i^1,v_i^1,s_i[1])\dots(u_i^n,v_i ^n,s_i[n]) 
    $$
    with each $(u_i^k,v_i^k,s_i[k])$ being an element of the set of generators $G\subseteq\ubu$, and such that these are the sequence of labels of two accepting runs in $\ese$. We'll see that these decompositions must be equal.

    We consider the first tuple where they differ: let's say it is the $k$-th. That is, $(u_1^k,v_1^k,s_1[k])\neq(u_2^k,v_2^k,s_2[k])$. There's two different ways they can differ:
    \begin{itemize}
        \item The empty word appears in a different place: that is $u_1^k = \emptyword \neq u_2^k$, or vice versa. At this point in the run, $\eme$ is in the same state for both computations. Since it is deterministic, the same state cannot read from two different tapes. Therefore, this option is impossible.
        \item The empty word appears in the same place, but the alphabet symbols are different: for instance if $u_1^k, u_2^k \neq \emptyword$ but $u_1^k\neq u_2^k$. This is also impossible, because both values must be equal to the same letter in $u$, as the same prefix of $u$ has been read already for both computations.
    \end{itemize}
    
    Thus, both decompositions must be equal, and so, $s_1=s_2$. Therefore $\ese$ is injective.

    We now show that $\ese$ is functional.

    Let $(u_1,v_1,s)$ and $(u_2,v_2,s)$ be two elements in $|\ese|$. We'll show that $(u_1,v_1)$ and $(u_2,v_2)$ must be equal. There must be two decompositions, for i=1,2:
$$
    (u_i,v_i,s)=(u_i^1,v_i^1,s[1])...(u_i^n,v_i^n,s[n]),
$$
    such that each tuple in the right is an element in $G$, and such that both are labels of successful runs in $\ese$. We will prove, by contradiction, that these two decompositions must be equal. Suppose they are different at some position k, that is:
$$
    (u_1^k,v_1^k,s[k])\neq(u_2^k,v_2^k,s[k])
$$  

    Because these are elements of $G$ and they are equal at the third coordinate, the only way they can be different is by $\emptyword$ appearing in different places. Since both decompositions are labels of successful computations in $\ese$, it means that at step k, the computations landed on different states (due to the determinism of $\ese$). Therefore, those two runs represent two different paths on the spliffer's graph.

    If we project the two obtained paths onto the underlying input automaton, we obtain two distinct paths with label $s$. This is a contradiction as the underlying input automaton is a classical deterministic automaton.
\end{proof}

\subsection{Closure Properties of Rational Sets}

Because $\ubu$ is finitely generated, the family $\rat(\ubu)$ of rational subsets of $\ubu$ corresponds exactly with the family of subsets accepted by spliffers. As it is the case with the set of rational subsets of any monoid, $\rat(\ubu)$ is closed under union, product and Kleene star. We are also interested in researching other closure properties of $\rat(\ubu)$, along with investigating the properties of $\drat(\ubu)$, the family of subsets of $\ubu$ accepted by deterministic spliffers. The most basic fact relating these two families is the following:

\begin{prop}
    $\drat(\ubu)$ is a proper subfamily of $\rat(\ubu)$.
\end{prop}

This is not hard to prove, as every deterministic spliffer is itself a spliffer, and not all behaviors of spliffers are injective functions. For instance, the behavior of the shuffler in Figure \ref{fig:elementary-shu-spl} is not injective, and therefore by Proposition \ref{behavior-is-injective-function} it can't be the behavior of a deterministic spliffer. The following table summarizes the closure results we have on these two sets:

\begin{center}
\begin{tabular}{ |r|c|c| }
 \hline
  & $\rat(\ubu)$ & $\drat(\ubu)$ \\
 \hline
 Union & Yes & No \\
 \hline
 Kleene Star & Yes & No \\ 
 \hline
 Product & Yes & No \\
 \hline 
 Intersection & No & No \\
 \hline
 Complementation & No & No \\
 \hline
 In $\drat(\ubu)$? & No & Trivial \\
 \hline
\end{tabular}
\end{center}

We now give counterexamples for the items marked with a "No" in the previous table.

\begin{lema}
    $\drat(\ubu)$ is not closed under union.
\end{lema}
\begin{proof}
    In general, the union of several injective functions needs not to be an injective function. For instance, $\conj{(a,b,ab)}$ and $\conj{(a,b,ba)}$ are separately in $\drat(\ubu)$ but its union is not.
\end{proof}

\begin{lema}
    Neither $\rat(\ubu)$ nor $\drat(\ubu)$ are closed under intersection. 
\end{lema}
\begin{proof}
    Consider the sets
    $$
    \ele_1 = \conj{\left(a^nba^m,a^mba^s,a^nba^{2m}ba^s \right):\;n,m,s>0} \\ 
    $$
    $$
    \ele_ 2 = \conj{\left(a^iba^j,a^kba^i,a^kba^{2i}ba^j\right):\; i,j,k>0}
    $$
    Both of these belong in $\drat(\ubu)$. For instance, Figure \ref{det-shu-spl-for-intersection} depicts a deterministic Spliffer whose behavior is $\ele_1$, and a similar one can be built for $\ele_2$. 

    \begin{figure}[h]
        \centering
        \begin{tikzpicture}[->,>=stealth',initial text=,semithick,auto,inner sep=3pt]
      \tikzstyle{every state}=[shape=circle split, minimum size=17];
      \node[state, style=state_left] (p0) at (0,0){};
      \node[state, style=state_right] (p1) at (3,0){};
      \node[state, style=state_left] (p2) at (3,-2){};
      \node[state, style=state_right] (p3) at (6,0){};
      
      \draw[<-] (p0.south) -- ++(-0.5,0cm);
      \draw[->, out=120, in=60, looseness=5] (p0.south west) to node[above]{$(a,\emptyword|a)$} (p0.north west);
      \path (p0) edge node {$(b,\emptyword|b)$} (p1);
      \path (p1) edge[bend left] node {$(\emptyword,a|a)$} (p2);
      \path (p2) edge[bend left] node {$(a,\emptyword|a)$} (p1);
      \path (p1) edge node {$(\emptyword,b|b)$} (p3);
      \draw[->, out=120, in=60, looseness=5] (p3.north east) to node[above]
      {$(\emptyword,a|a)$} (p3.south east);
      \draw[->] (p3.south) -- ++(0.5,0cm);
    \end{tikzpicture}
    \caption{}\label{det-shu-spl-for-intersection}

    \end{figure}

    However, the intersection between them:
    $$
    \ele_1\cap\ele_2 = \conj{\left(a^sba^s,a^sba^s,a^sba^{2s}ba^s\right):\; s>0}
    $$
    is not accepted by \textit{any} spliffer, deterministic or not, because its projection onto the third coordinate is not a regular language, as it would be required by Corollary \ref{projection-is-regular} if it were. In particular, this example proves that neither $\rat(\ubu)$ nor $\drat(\ubu)$ are closed under intersections.
    
\end{proof}

\begin{coro}
    $\rat(\ubu)$ is not closed under complementation.
\end{coro}
\begin{proof}
    Because it is closed under union but not intersection.
\end{proof}

\begin{lema}
    $\drat(\ubu)$ is not closed under complementation, product nor Kleene star.
\end{lema}
\begin{proof}
    In general, the complement of an injective function may not be an injection nor a function. For instance, consider the set:
    $$
    \ele =  \conj{(b,a^n,ba^n):n\in\enne}
    $$
    which is the behavior of the deterministic spliffer depicted in Figure \ref{det-shu-spl-for-complement}.
    
    \begin{figure}[h]
        \centering
        \begin{tikzpicture}[->,>=stealth',initial text=,semithick,auto,inner sep=3pt]
      \tikzstyle{every state}=[shape=circle split, minimum size=17];
      \node[state, style=state_left] (p0) at (0,0){};
      \node[state, style=state_right] (p3) at (3,0){};
      
      \draw[<-] (p0.south) -- ++(-0.5,0cm);
      \path (p0) edge node {$(b,\emptyword|b)$} (p3);
      \draw[->, out=120, in=60, looseness=5] (p3.north east) to node[above]{$(\emptyword,a|a)$} (p3.south east);
      \draw[->] (p3.south) -- ++(0.5,0cm);
    \end{tikzpicture}
    \caption{}\label{det-shu-spl-for-complement}
    \end{figure}
    
    Among other things, the complement of $\ele$ with respect to $\ubu$ contains the elements $(b,a^n,a^nb)$ and $(b,a^n,a^{n-1}ba)$ for all $n\geq2$. In particular, $\ubu\setminus\ele$ cannot be the behavior of a deterministic splitter, as it is not an injection. We conclude that $\drat(\ubu)$ is not closed under complementation.

    Additionally, both $\ele^\star$ and $\ele\ele$ contain simultaneously elements of the form $(bb,a^{n+m},ba^nba^m)$ and $(bb,a^{n+m},ba^mba^n)$ for different $n,m\geq1$. In particular, neither $\ele^\star$ nor $\ele\ele$ can be the behavior of a deterministic splitter, because they are not injections. So we are also forced to conclude that $\drat(\ubu)$ is not closed under product nor Kleene star. 
\end{proof}

\section{A Theorem on Spliffer Equivalence}\label{spliffer-equiv}

The main result of this Section is the following Theorem.

\begin{teo}\label{teo:decidable_equivalence}
    The equivalence between two functional splitters is decidable. 
\end{teo}

We prove Theorem \ref{teo:decidable_equivalence} by adapting a technique developed by Béal, Carton, Prieur and Sakarovitch in \cite{beal2003squaring} to prove the decidability of equivalence of functional transducers. The proof makes use of the following ingredients (out of which the second is the hardest to obtain):

\begin{enumerate}
    \item It's decidable whether two functional splitters share their domain.
    \item It's decidable whether a given splitter is functional.
    \item It's possible to build a splitter which accepts the union of the behaviors of two different splitters.
\end{enumerate}

Point number $1$ amounts to decide whether two classical automata over $A^\star$ are equivalent, because the domain of a splitter is simply the behavior of its underlying input automaton. Point number $3$ is an elementary property of the family $\rat(\ubu)$. We now concern ourselves with point number $2$ and, for that, we first recall a few definitions. References are taken from \cite{sakarovitch2009elements}.

\subsection{Valuations and the Lead or Delay Action}

Given a right action $\delta:X\times N\rightarrow X$ of a monoid $N$ over the set $X$, with distinguished element $x_0\in X$, and an automaton $\abu=\autom$ over $N$ we can always consider the product automaton $\abu\times\delta$ which executes the transitions of the automaton $\abu$ and performs the action $\delta$ at the same time. This automaton is defined as the trimmed part of $(Q\times X,M,E',I\times \conj{x_0},T\times X)$, where $(q,x)\xrightarrow{m}(q',x')$ if and only if $q\xrightarrow{m}q'$ and $x' = x\cdot_\delta m$.

We call the automaton $\abu\times\delta$ a \textit{valuation} of $\abu$ whenever the projection $\pi:\abu\times\delta\rightarrow\abu$ is a bijection. In short, the product with an action is a valuation whenever the action assigns to each state of $\abu$ a single value of its acting set $X$. We can therefore talk about the \textit{values} of states $q\in Q$ and denote them as $v(q)$.

In our case, we start with a splitter $\ese$ and consider a valuation over the \textit{square automaton} $\ese\times\ese = (Q\times Q, A, \tilde{E}, I\times I, T\times T)$ with transitions
$$
(q',q'') \xrightarrow[\ese\times\ese]{(a',a'',a)} (r',r'')
$$
whenever $a\in A$, $(a',a),(a'',a)\in G$ and $q'\xrightarrow{(a',a)}r'$, $q''\xrightarrow{(a'',a)}r''$ are two transitions in $\ese$.

Intuitively, $\ese\times\ese$ is an automaton which executes the transitions of two copies of $\ese$ simultaneously. Note that it is no longer a splitter, as it can be thought to have four output tapes instead of two. A possible label for a transition in $\ese\times\ese$, for example, is $((a,\emptyword),(\emptyword,a),a)$. Here, the third coordinate represents the automaton's input, the first coordinate represents the output of the first copy of $\ese$ and the second coordinate the output of the second copy of $\ese$.  

In the context of deciding functionality, it makes sense to study a construct like this, as it is a specimen to compare the output of different runs of $\ese$ with the same input. To be able to effectively run this comparison, we make use of an action $\delta$ called the \textit{Lead or Delay Action}. 

The Lead or Delay Action is an action by the monoid $A^\star\times A^\star$ over a set $H_A$ which, informally, keeps count of how far ahead is one word with respect to another. We define

$$
H_A = \left(A^\star\times\conj{\emptyword}\right) \cup \left(\conj{\emptyword}\times A^\star\right) \cup \conj{\bold 0}
$$

The element $(f,\emptyword)\in H_A$ represents a \say{lead} $f$ of the first word with respect to the second, meaning it is a representative of tuples of words with the form $(wf,w)$. On the contrary, an element $(\emptyword,g)$ represents tuples with the form $(w,wg)$, in other words, a \say{delay} $g$ of the first word with respect to the second. The pairs of words that have nothing to do with each other, meaning neither is a prefix of the other, are represented with the element $\bold0$. We note $w\leq w'$ to mean that $w$ has a \say{delay} with respect to $w'$, which in usual terms it simply means that $w$ is a prefix of $w'$. Also, whenever $f\leq g$ and $g=fw$, we note $f^{-1}g = w$.

The \textit{Lead or Delay Action} is a right action $\delta:H_A\times(A^\star\times A^\star)\rightarrow H_A$ that updates the advancement or delay of the word on the first tape of a transducer with respect to the word on the second after concatenating with the pair of words in $A^\star\times A^\star$. Concretely, when $(h_l,h_r)\neq\bold0$ we define it as:

$$
(h_l,h_r)\cdot_\delta(f,g) = \begin{cases} ((h_rg)^{-1}h_lf,\emptyword) &\text{ if } h_rg\leq h_lf \\ (\emptyword,(h_lf)^{-1}h_rg) &\text{ if } h_lf\leq h_rg \\ \bold0 &\text{ otherwise.}\end{cases}
$$  
and $\bold0\cdot_\delta(f,g) = \bold0$ for every pair of words $(f,g)\in A^\star\times A^\star$. 

However, it's not clear from this definition that $\delta$ is an action at all. Clearing that up is the first thing that needs to be done.

\begin{lema}
    The Lead or Delay Action is a right action by the monoid $A^\star\times A^\star$ over the set $H_A$.
\end{lema}

\begin{proof}
    Take $(h_l,h_r)\in H_A$ and $f,f',g,g'\in A^\star.$ We show that
    
    $$(h_l,h_r)\cdot_\delta(ff',gg')=((h_l,h_r)\cdot_\delta~(f,g))\cdot_\delta(f',g').$$ 
    
    The right hand side reduces to
\begin{align*}
    ((h_l,h_r)\cdot_\delta(f,g))\cdot_\delta(f',g') &= \begin{cases} 
    (\emptyword,(h_lf)^{-1}h_rg)\cdot_\delta(f',g') & \text{ if } h_lf\leq h_rg \\
    ((h_rg)^{-1}h_lf,\emptyword)\cdot_\delta(f',g') & \text{ if } h_rg\leq h_lf \\
    \bold0 & \text{ otherwise.}
    \end{cases} \\
    &= \begin{cases}
    (\emptyword,(h_lff')^{-1}h_rgg') & \text{ if } h_lf\leq h_rg \text{ and } f'\leq(h_lf)^{-1}h_rgg' \\
    (((h_lf)^{-1}h_rgg')^{-1}f',\emptyword) & \text{ if } h_lf\leq h_rg \text{ and } (h_lf)^{-1}h_rgg' \leq f' \\
    ((h_rgg')^{-1}h_lff',\emptyword) & \text{ if } h_rf\leq h_lg \text{ and } g'\leq(h_lg)^{-1}h_lff' \\
    (\emptyword,((h_rg)^{-1}h_lff')^{-1}g') & \text{ if } h_rg\leq h_lf \text{ and } (h_rg)^{-1}h_lff'\leq g'\\
    \bold 0 & \text{ any other case.}
    \end{cases}
\end{align*}

However, it is a common sense property that for any choice of words $u,v,w$ such that $u\leq v$ and $u^{-1}v\leq w$, then $((u^{-1}v)^{-1}w = v^{-1}uw$. This implies that cases $1$ and $4$, and cases $2$ and $3$ in the previous equation reduce to the same element of $H_A$. Then

$$
    ((h_l,h_r)\cdot_\delta(f,g))\cdot_\delta(f',g') = 
    \begin{cases}
    (\emptyword,(h_lff')^{-1}h_rgg') & \text{ if } (h_lf\leq h_rg \text{ and } f'\leq(h_lf)^{-1}h_rgg')  \\
    & \hspace{3mm}\text{ or } (h_rg\leq h_lf \text{ and } (h_rg)^{-1}h_lff'\leq g')\\
    ((h_rgg')^{-1}h_lff',\emptyword) & \text{ if } (h_lf\leq h_rg \text{ and } (h_lf)^{-1}h_rgg' \leq f') \\
    & \hspace{3mm}\text{ or } (h_rf\leq h_lg \text{ and } g'\leq(h_lg)^{-1}h_lff') \\
    \bold 0 & \text{ any other case.}
    \end{cases}
$$

This is exactly the same as $(h_l,h_r)\cdot_\delta(ff',gg')$, because the first clause is equivalent to $h_lff'\leq h_rgg'$ and the second is equivalent to $h_rgg'\leq h_lff'$.

\end{proof}

The Lead or Delay Action action satisfies two properties that we need:
\begin{lema}
\begin{itemize}
\item[]
    \item The stabilizer of $(\emptyword,\emptyword)$ under $\delta$ is the set $\{(w,w):w\in A^\star\}$.
    \item Semi-injectivity: if $X\cdot_\delta(f,g) = Y\cdot_\delta(f,g)$ and both sides are different from $\bold0$, then $X=Y$.
\end{itemize}
\end{lema}

\begin{proof}
    The first property is immediate from definition. We prove here the property of semi-injectivity. Suppose that $X\cdot_\delta(f,g) = Y\cdot_\delta(f,g) = Z$ with $X,Y,Z\in H_A\setminus\conj{\bold0}$ and $f,g\in A^\star.$ Without loss of generality, suppose that $Z=(z,\emptyword)$ with $z\in A^\star$. We first show that $X$ is almost unequivocally determined from the equation $X\cdot_\delta (f,g)=(z,\emptyword)$. For this, we consider two distinct cases.

\begin{enumerate}
    \item First, assume that $X=(x,\emptyword)$ with $x\in A^\star$. The fact $(x,\emptyword)\cdot_\delta(f,g)=(z,\emptyword)$ implies $g\leq xf$ and $z = g^{-1}xf$. In particular, $gz = xf$. Then, the word $gz$ ends in $f$ and to obtain $x$ it suffices to remove $f$ from the end of $gz$. Thus, $x$ is determined from $f,g$ and $z$, and therefore $X=(x,\emptyword)$ is determined from them as well. This case is illustrated in the following figure.

\begin{center}
    \begin{tikzpicture}

    \fill[gray!20] (3.35,0.75) rectangle (7,1.25);
    
    \draw (0,0) rectangle (7,0.5);
    \draw (0,0.75) rectangle (7,1.25);

    \draw (2,0) -- (2,0.5);
    \draw (3.35,0.75) -- (3.35,1.25);
    
    \node at (1,0.25) {x};
    \node at (4.5,0.25) {f};
    \node at (1.8,1) {g};
    \node at (5,1) {z};

    \end{tikzpicture}
\end{center}

    \item Assume now that $X=(\emptyword,x)$ with $x\in A^\star$. From $(\emptyword, x)\cdot_\delta(f,g)=(z,\emptyword)$ it can be deduced that $xg\leq f$ and $z = (xg)^{-1}f$. Then, $xgz = f$, and to obtain $x$ it suffices to remove the word $gz$ from the end of $f$. Therefore, in this case $x$ can also be obtained unequivocally from $f,g$ and $z$. This is illustrated in the following figure.

\begin{center}
    \begin{tikzpicture}

    \fill[gray!20] (3.6,0.75) rectangle (7,1.25);
    
    \draw (0,0) rectangle (7,0.5);
    \draw (0,0.75) rectangle (7,1.25);

    \draw (2,0.75) -- (2,1.25);
    \draw (3.6,0.75) -- (3.6,1.25);
    
    \node at (3.5,0.25) {f};
    \node at (1,1) {x};
    \node at (2.8,1) {g};
    \node at (5.3,1) {z};

    \end{tikzpicture}
\end{center}

\end{enumerate}

    It remains to be shown that there can't be an $X=(x,\emptyword)$ and a $Y=(\emptyword,y)$ (with the empty word in different places) such that $X\cdot_\delta(f,g)=Y\cdot_\delta(f,g) = (z,\emptyword)$. Assume that such $X,Y$ exist. Then, by case $1$ it follows that $xf = gz$ and by case $2$ it follows that $ygz = f$. Replacing the second equation into the first, we obtain $xygz=gz$. So both $x,y$ must be equal to $\emptyword$.
\end{proof}

However, in our particular case we are aiming to compare the outputs of two splitters. That is, we'll have two sets of two words to compare, rather than a single set. That's why we define the \textit{bidimensional Lead or Delay} action, which consists in applying the Lead or Delay action twice. We define $\Delta:(H_A\times H_A)\times ((A^\star\times A^\star)\times(A^\star\times A^\star))\rightarrow(H_A\times H_A)$ as follows:
$$
(h',h'')\cdot_\Delta((f',g'),(f'',g'')) = (h'\cdot_\delta(f',f''),h''\cdot_\delta(g',g''))
$$

Here $(f',g')$ represents the output of the first copy of $\ese$, and $(f'',g'')$ the output of the second copy. Thus, $(f',f'')$ contains the left outputs of the two copies of $S$, and $(g',g'')$ contains the right outputs. $\Delta$ can thus be understood to be comparing the advancement of both copies of $\ese$ tape by tape. 

The following properties of $\Delta$ are inherited from $\delta$, and their proofs are straightforward.
\begin{lema}\label{lemma:semi-injectivity}
\begin{itemize}
    \item[]
    \item $\Delta$ is a right action by the monoid $(A^\star\times A^\star)\times(A^\star\times A^\star)$ over the set $H_A\times H_A$.
    \item The stabilizer of $\emptywords$ under $\Delta$ is the set $\conj{((v,w),(v,w)):v,w\in A^\star}$.
    \item Semi-injectivity: if $X\cdot_\Delta((f', g'), (f'',g'') = Y\cdot_\Delta((f', g'), (f'',g'')$ and no $\bold0$ appears in the left or right sides, then $X=Y$.
\end{itemize}
\end{lema}

\subsection{The Equivalence of Functional Splitters}

To decide the functionality of a given splitter $\ese$, we consider the product between the square automaton and the bidimensional Lead or Delay action, that is $(\ese\times\ese)\times\Delta$. We make a slight abuse of notation to ensure everything works as it should: $\Delta$'s acting monoid is extended to $((A^\star\times A^\star)\times(A^\star\times A^\star)\times A^\star)$ but the third coordinate is ignored. That is $(h',  h'') \cdot_\Delta (a',a'',a) = (h',  h'') \cdot_\Delta (a', a'')$. The main result is as follows:

\begin{teo}\label{functionality-is-decidable}
    Let $\ese$ be an splitter, and let $\abu$ be the trim part of its square automaton $\ese\times\ese$. Let $\Delta$ be the bidimensional Lead or Delay Action. Then $\ese$ is functional if and only if the two following conditions hold:
    \begin{itemize}
        \item $\abu\times\Delta$ is a valuation of $\abu$.
        \item The value $v(q',q'')$ of each final state $(q',q'')$ of $\abu$ under $\Delta$ is $\emptywords$. 
    \end{itemize}
\end{teo}

\begin{proof}
    We start by showing the sufficiency of those two conditions. We consider two successful runs in the splitter $\ese$ with the same input $a_1\dots a_n\in A^\star$, say

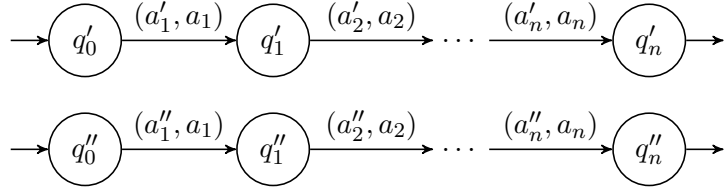
\begin{center}
    \begin{tikzpicture}[->,>=stealth',initial text=,semithick,auto,inner sep=3pt, node distance=2.5cm]
    \node[state, initial] (q0) at (0,0) {$q'_0$};
    \node[state, right of=q0] (q1) {$q'_1$};
    \node[right of=q1] (dots) {$\dots$};
    \node[state, accepting right, right of=dots] (qn) {$q'_n$};
  
    \path[->]
        (q0) edge node [above] {$(a'_1,a_1)$} (q1)
        (q1) edge node [above] {$(a'_2,a_2)$} (dots)
        (dots) edge node [above] {$(a'_n,a_n)$} (qn);
    \end{tikzpicture}
\end{center}
\begin{center}
    \begin{tikzpicture}[->,>=stealth',initial text=,semithick,auto,inner sep=3pt, node distance=2.5cm]

    \node[state, initial] (r0) at (0,-1.5) {$q''_0$};
    \node[state, right of=r0] (r1) {$q''_1$};
    \node[right of=r1] (rots) {$\dots$};
    \node[state, accepting right, right of=rots] (rn) {$q''_n$};
  
    \path[->]
        (r0) edge node [above] {$(a''_1,a_1)$} (r1)
        (r1) edge node [above] {$(a''_2,a_2)$} (rots)
        (rots) edge node [above] {$(a''_n,a_n)$} (rn);
    \end{tikzpicture}
\end{center}
where each $a'_i,a''_i$ might be either $(\emptyword,a_i)$ or $(a_i,\emptyword)$.

Both the initial and the final states of $\ese\times\ese$ have value $\emptywords$ under the valuation induced by $\Delta$. Therefore:
\begin{align*}
\emptywords &= v(q'_n,q''_n) \\&= v(q'_0,q''_0)\cdot_\Delta (a'_1\dots a'_n, a''_1\dots a''_n)\\ &= \emptywords \cdot_\Delta (a'_1\dots a'_n, a''_1\dots a''_n)
\end{align*}
We conclude that the element on the right hand of the equation must lie in the stabilizer of $\emptywords$, and so $a'_1\dots a'_n = a''_1\dots a''_n$.

We now prove the necessity of those conditions. Let's first suppose that $\Delta$ effectively induces a valuation on $\ese$, but at least one final state $(t',t'')$ has a value different from $\emptywords$. Let $(w',w'',w)$ be the label of a successful run of $\abu$ ending in $(t',t'')$. In particular,
$$
\emptywords\cdot_\Delta(w',w'') \neq \emptywords. 
$$
So $(w',w'')$ does not belong to the stabilizer of $\emptywords$. Therefore, $w'\neq w''$, and $\ese$ is not functional.

Finally, suppose that $\abu\times\Delta$ isn't even a valuation. Then, there must exist a state $(p',p'')$ of $\abu$ with more than one possible assigned value. It follows that it must be possible to reach that state in two different ways. Because $\abu\times\Delta$ is trim, any computation that ends in $(p',p'')$ can be completed to a successful run. We can therefore draw the following diagram, which depicts two different successful runs passing through $(p',p'')$:

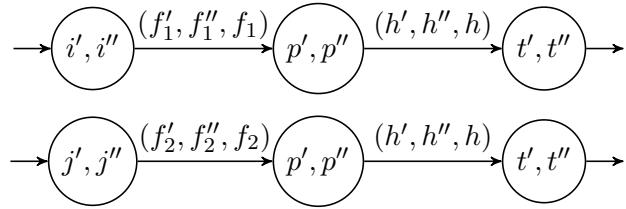
    \begin{figure}[h]
        \centering
    \begin{tikzpicture}[->,>=stealth',initial text=,semithick,auto,inner sep=3pt, node distance=3cm]
    \node[state, initial] (q0) at (0,0) {$i',i''$};
    \node[state, right of=q0] (q1) {$p',p''$};
    \node[state, accepting right, right of=q1] (q2) {$t',t''$};

    \node[state, initial] (r0) at (0,-1.5) {$j',j''$};
    \node[state, right of=r0] (r1) {$p',p''$};
    \node[state, accepting right, right of=r1] (r2) {$t',t''$};
  
    \path[->]
        (q0) edge node [above] {$(f'_1,f''_1,f_1)$} (q1)
        (q1) edge node [above] {$(h',h'',h)$} (q2)
        (r0) edge node [above] {$(f'_2,f''_2,f_2)$} (r1)
        (r1) edge node [above] {$(h',h'',h)$} (r2);
    \end{tikzpicture}
\end{figure}
where 
$$
\emptywords \cdot_\Delta (f'_1, f''_1) \neq \emptywords \cdot_\Delta (f'_2, f''_2)
$$

We can apply $(h',h'')$ to both sides of the previous inequality. If any of those applications result in any $\bold{0}$, say the left one, that'd mean that $\ese$ is not functional because the input $f_1h$ results in at least two different outputs. On the other hand, if no $\bold{0}$ appears as a result from the application, then by Lemma \ref{lemma:semi-injectivity} we obtain two different values under $\Delta$ for the final state $(t',t'')$.

Then, one of these values is not $\emptywords$. Following a similar reasoning as in the previous case, we conclude that $\ese$ is not functional.  

\end{proof}

\begin{coro} \label{coro:deciding_functionality_is_quadratic}
    Let $\ese$ be a splitter with states set $Q_\ese$. The functionality of $\ese$ can be decided in quadratic time over $|Q_\ese|$.
\end{coro}

\begin{proof}
    We can do the following:
\begin{enumerate}
    \item Build $\abu = \ese\times\ese$ and trim it. Its states set $Q_\abu$ is a subset of $Q_\ese\times Q_\ese$. 
    \item Build $\abu\times\Delta$. This automaton could potentially have an infinite number of accessible states, so end in failure if at some point there's more than $|Q_\abu|$ states built.
    \item Check whether $\abu\times\Delta$ is a valuation and all of its final states are valued $\emptywords$.
\end{enumerate}

$\abu$ has at most $|Q_\ese|^2$ states, so it takes roughly quadratic time to build it on step $1$. Step $2$ runs, at the most, for as much time as $\abu$ has states: if more states of $\abu\times\Delta$ are built, then it's not a valuation. Step $3$ requires checking all states of $\abu\times\Delta$ to verify that it is actually a valuation and all its final states have the desired value.  
\end{proof}

With this, we can prove the main result.

\begin{proof}[Proof of Theorem \ref{teo:decidable_equivalence}]
Given functional splitters $\ese_1$ and $\ese_2$, we do the following:
\begin{enumerate}
    \item Verify whether $\ese_1$ and $\ese_2$ share the same domain. If they don't, they are trivially not equivalent.
    \item Build a splitter $\ese$ such that $|\ese| = |\ese_1|\cup|\ese_2|$.
    \item Verify whether $\ese$ is itself functional as a splitter. If it is, then it follows that $|\ese_1|=|\ese_2|$ because those two functions share its domain. If it isn't, then $\ese_1$ and $\ese_2$ are not equivalent.
\end{enumerate}
Step $1$ can be performed with elementary techniques from classical automata theory. Step $2$ amounts to non-deterministically run either $\ese_1$ or $\ese_2$. Step $3$ is an application of the previous technique. 
\end{proof}

Note that this approach requires regarding the spliffers exclusively as splitters. One can also define shufflers to be functional in the natural way, and with this definition deterministic spliffers are also functional shufflers. However, an algorithm similar to the one in Theorem \ref{teo:decidable_equivalence} applied to functional shufflers fails. For instance, point $1$ of the algorithm would require to verify the equivalence between two arbitrary two-tape transducers, something which is known to be undecidable \cite{fischer1968multitape}. 

Theorem \ref{teo:decidable_equivalence} has as an immediate corollary the decidability of equivalence between deterministic spliffers. Also, in the deterministic case we can ensure polynomial times for the procedure.

\begin{coro}
    If $\ese_1$ and $\ese_2$ are deterministic spliffers with state sets $Q_1$ and $Q_2$ respectively, their equivalence can be decided in quadratic time over $|Q_1|+|Q_2|$. 
\end{coro}
\begin{proof}
    Deterministic spliffers are functional splitters by Proposition \ref{behavior-is-injective-function}, so the previous algorithm can be used to determine the equivalence between $\ese_1$ and $\ese_2$. Because the splitters are deterministic, their underlying input automata are classical deterministic automata, so their equivalence can be decided in less than quadratic time, for instance by minimizing them and checking for equality (see \cite{hopcroft1971n}, for example, for an efficient minimization algorithm). Step $2$ of the algorithm is linear on $|Q_1|+|Q_2|$. Step $3$ of the algorithm can be performed in quadratic time because of Corollary \ref{coro:deciding_functionality_is_quadratic}.
\end{proof}

\section*{Acknowledgements}

The research that led to this paper was funded by grant PICT 2018-2315 of Agencia Nacional de Promoción Científica y Tecnológica de Argentina.
I acknowledge my supervisors, Prof. Verónica Becher and Prof. Martin Mereb, as well as Prof. Nicolás Álvarez, for the discussions we have every Thursday, which led to the results presented here.
I also thank Prof. Olivier Carton for the discussions we had and his suggestion to look into functional transducers, and also to Prof. Jacques Sakarovitch for kindly agreeing to look at an earlier version of this document, and providing useful feedback.

\bibliographystyle{ieeetr}
\bibliography{sample}

\begin{thebibliography}{1}

\bibitem{alvarez2019finite}
Nicol{\'a}s Alvarez, Ver{\'o}nica Becher, and Olivier Carton.
\newblock Finite-state independence and normal sequences.
\newblock {\em Journal of Computer and System Sciences}, 103:1--17, 2019.

\bibitem{beal2003squaring}
Marie-Pierre B{\'e}al, Olivier Carton, Christophe Prieur, and Jacques
  Sakarovitch.
\newblock Squaring transducers: an efficient procedure for deciding
  functionality and sequentiality.
\newblock {\em Theoretical Computer Science}, 292(1):45--63, 2003.

\bibitem{boasson2015rational}
Luc Boasson and Olivier Carton.
\newblock Rational selecting relations and selectors.
\newblock In {\em International Conference on Language and Automata Theory and
  Applications}, pages 716--726. Springer, 2015.

\bibitem{fischer1968multitape}
Patrick~C Fischer and Arnold~L Rosenberg.
\newblock Multitape one-way nonwriting automata.
\newblock {\em Journal of Computer and System Sciences}, 2(1):88--101, 1968.

\bibitem{harju1991equivalence}
Tero Harju and Juhani Karhum{\"a}ki.
\newblock The equivalence problem of multitape finite automata.
\newblock {\em Theoretical Computer Science}, 78(2):347--355, 1991.

\bibitem{hopcroft1971n}
John Hopcroft.
\newblock An n log n algorithm for minimizing states in a finite automaton.
\newblock In {\em Theory of machines and computations}, pages 189--196.
  Elsevier, 1971.

\bibitem{sakarovitch2009elements}
Jacques Sakarovitch.
\newblock {\em Elements of automata theory}.
\newblock Cambridge University Press, 2009.

\end{thebibliography}

\end{document}